\documentclass[twocolumn,secnumarabic,amssymb, nobibnotes, aps, prd,superscriptaddress]{revtex4-2}
\usepackage{graphicx}
\usepackage{amsmath}
\usepackage{booktabs}

\begin{document}

\title{Machine learning-based direct solver for one-to-many problems on temporal shaping of relativistic electron beams}%

\author{Jinyu Wan}%
\affiliation{Key Laboratory of Particle Acceleration Physics and Technology, Institute of High Energy Physics, Chinese Academy of Sciences, Beijing 100049, China}
\affiliation{University of Chinese Academy of Sciences, Beijing 100049, China}
\affiliation{China Spallation Neutron Source, Dongguan Guangdong 523803, China}
\author{Yi Jiao}%
\email[Corresponding author:]{\\jiaoyi@ihep.ac.cn}
\affiliation{Key Laboratory of Particle Acceleration Physics and Technology, Institute of High Energy Physics, Chinese Academy of Sciences, Beijing 100049, China}
\affiliation{University of Chinese Academy of Sciences, Beijing 100049, China}
\date{\today}%

\begin{abstract}
	To control the temporal profile of a relativistic electron beam to meet requirements of various advanced scientific applications like free-electron-laser and plasma wakefield acceleration, a widely-used technique is to manipulate the dispersion terms which turns out to be one-to-many problems. Due to their intrinsic one-to-many property, current popular stochastic optimization approaches on temporal shaping may face the problems of long computing time or sometimes suggesting only one solution. Here we propose a real-time solver for one-to-many problems of temporal shaping, with the aid of a semi-supervised machine learning method, the conditional generative adversarial network (CGAN). We demonstrate that the CGAN solver can learn the one-to-many dynamics and is able to accurately and quickly predict the required dispersion terms for different custom temporal profiles. This machine learning-based solver is expected to have the potential for wide applications to one-to-many problems in other scientific fields.
\end{abstract}
\maketitle
\section{Introduction}\label{I}
Much of the interest in advanced scientific applications of particle accelerator user facilities such as free-electron lasers (FELs) \cite{ref1,ref2,ref3}, terahertz radiation \cite{ref4,ref5} and plasma wakefield accelerators (PWFAs) \cite{ref6,ref7} has grown in the past decades. FELs and terahertz radiation are emerging as powerful imaging tools in various fields \cite{ref8,ref9,ref10,ref11} like physics, chemistry, biology and material science. PWFAs are potential to provide an accelerating gradient up to multi-GV/m level for high energy physics and photon science \cite{ref12,ref13}. Such a gradient is much higher than that obtained with conventional radio-frequency-based accelerators. To realize these advanced accelerator applications, a critical issue is to provide relativistic electron beams of particular temporal shapes to improve quality of the electron or photon beam, namely temporal shaping of electron beams \cite{ref14,ref15,ref16,ref17,ref18,ref19}. For example, a linearly ramped beam is required by PWFAs to supply a high transformer ratio of wakefield \cite{ref14,ref15}, and beams with flat-top temporal profiles are desired in FELs to obtain high radiation performance \cite{ref16,ref17}.\par
One widely-used temporal shaping method is to let an electron bunch with inhomogeneous energy distribution pass through an energy dispersion section, like a so-called bunch compressor consisting of several bending magnets to realize various temporal profiles \cite{ref20}. For beams with small transverse emittances and large energy spread, the contribution of geometrical terms relating to the transverse emittances can be neglected in the transfer map and the dispersion terms that connect to the energy spread will tend to dominate the process \cite{ref21}. Hence, for an initial beam having an inhomogeneous energy chirp, a desired custom temporal profile can be achieved by finding an appropriate combination of dispersion terms (see details in Methods section). Due to the highly nonlinear process, different combinations of dispersion terms may realize the same target profile, i.e. the temporal shaping can intrinsically turn into a one-to-many problem. The exploration of multiple solutions instead of only one is beneficial for various scientific applications, e.g. for beams with the same temporal profile, but with opposite energy chirp along the electron beam can provide different benefits in FELs \cite{ref22,ref23}. In addition, it should be noted that not all potential solutions are equally feasible to implement in practical scientific experiments, e.g. one may require unfeasibly stronger magnets than those in other solutions. Therefore, if possible, it is beneficial to derive multiple, if not all, potential solutions that can result in almost the same target profile.\par
Grid scan \cite{ref24} was first used to solve such a temporal shaping problem to realize a ramped profile with a second-order approximation. However, due to the highly nonlinear process, the second-order approximation would be insufficient. Yet, when contributions of the higher-order longitudinal dispersion terms are included in the scan, the time cost of grid scan can grow exponentially. Later, it has been shown \cite{ref25,ref26,ref27,ref28} that such one-to-many problems can be solved more efficiently with stochastic optimization methods like genetic algorithm (GA) \cite{ref29,ref30}, particle swarm optimization (PSO) \cite{ref31} and extremum seeking (ES) \cite{ref32}. Nevertheless, the stochastic optimization process is still indirect and some challenging problems still remain. For a single-objective optimization problem that have multiple potential solutions, the optimization process sometimes stops when the first solution is found if insufficiently large population size is given, resulting in the omission of other potential solutions that may be more feasible. Although one can implement additional constrains to see whether an obtained solution is feasible or not, the constrained optimization can be more difficult to solve and more computationally expensive \cite{ref33}. In this case, the finally obtained results can be highly dependent on the choice of initial solutions. So a warm start is critical for most of these optimization algorithms. \par
In recent years, machine learning (ML) has attracted increasing interests of accelerator experts as a powerful tool to reveal the complicated correlations between various accelerator parameters \cite{ref34,ref35,ref36,ref37,ref38,ref39,ref40,ref41,ref42}. It is noticed that, however, most of these ML applications are based on supervised ML models, which are only powerful to capture the map of one-to-one problems where one feature vector \textit{X} has only one definite label vector \textit{Y}. When a supervised ML model is trained with data samples of a one-to-many problem that has different labels for the same input feature vector, it tends to output the mean label value of the samples rather than the respective label value of each sample itself. For example, a supervised ML model is able to predict the temporal profile of an electron bunch with known accelerator settings \cite{ref36}. While for the inverse problem, namely, predicting the accelerator settings for a desired custom temporal profile, the supervised ML methods like multilayer perceptron may fail to give the right answer because multiple solutions possibly exist.\par
In this paper, we propose a direct solver for one-to-many problems of beam profile shaping with the aid of a special ML method, the conditional generative adversarial networks (CGAN) \cite{ref43}.  The generative adversarial networks (GANs) \cite{ref44} are emerging techniques of unsupervised and semisupervised ML that has potential to handle one-to-many problems \cite{ref45,ref46}. Until now, the GANs have become one of the state-of-the-art techniques to solve difficulties in image synthesis \cite{ref47}, style transfer \cite{ref48}, and image superresolution \cite{ref49}. The purpose of GANs is to train a generative model to discover and learn the regularities or patterns in training data so that the model can be used to generate new samples that appear to be drawn from the training dataset. Instead of minimizing the mean error as in most supervised ML methods, GANs take a different approach to learn correlations hidden in the training data, via the competition of a pair of neural networks, the generator and the discriminator. The generator is trained to create fake data samples as authentic as possible to fool the discriminator. The discriminator is trained to output the probability of whether an input sample is from the original training dataset or is generated by the generator. The two neural networks are trained together in a zero-sum game to be adversarial. Successful training of a GAN usually ends with the discriminator outputting the same probability for both real and fake samples, meaning the generator model is generating plausible examples. After training, the fixed-length random vector is randomly drawn from a Gaussian distribution and is used to seed the generative process. A random vector will correspond to points in the problem domain, forming a compressed representation of the data distribution. For one-to-many problems, each solution can correspond to different input vector, such that a new random input vector provided to the generator model as input can generate new and different output samples. \par
CGAN is an important extension to the GANs. Compared to GANs that generate new samples from a random input noise vector, a CGAN can generate samples of a given type by making both the generator and the discriminator class-conditional. With an additional label added to the input data, both the generator and the discriminator receive additional information about the correlation of the samples and the given label so that the networks can synthesize samples with user specified content. This extended approach allowing one to direct the data generation process has proven to be effective to create images with a target class \cite{ref50,ref51}. The power of CGAN to generate data samples for a given class implies the potential of solving inverse design problems like temporal profile shaping of electron beams. \par
The work is structured as follows. In Section 2, we review the temporal shaping problem and describe the CGAN solver. In Section 3, we present the results of using CGAN solver to realize two typical temporal profiles and two other temporal profiles with more scientific merits. The application of the CGAN solver to more complicated situation with CSR effect considered is also investigated. Finally, conclusions are summarized in Section 4.\par

\begin{figure*}[htbp]    
	\vspace*{1mm}    
	\centering    
	\includegraphics[scale=0.3]{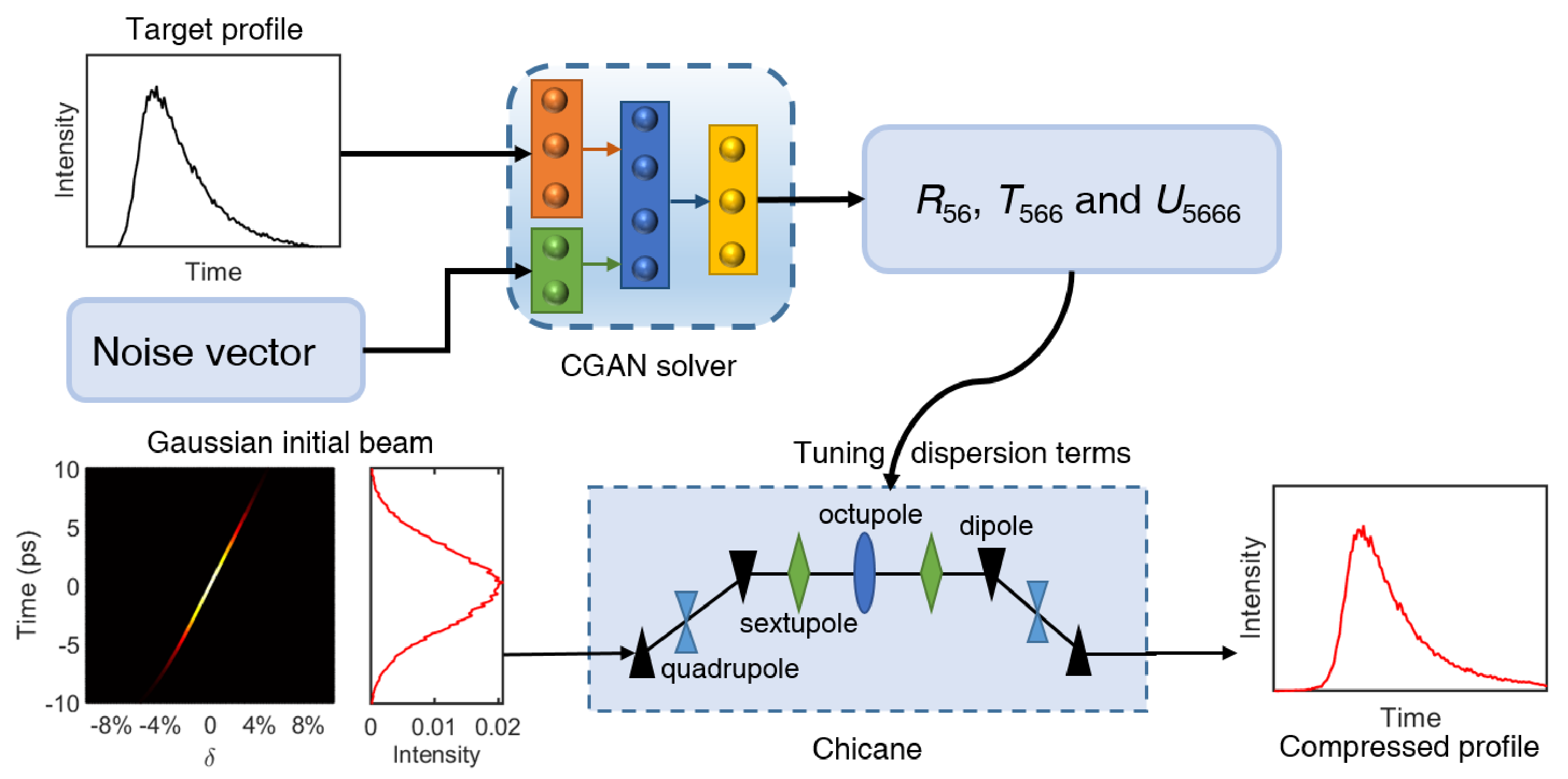}    
	\\[1mm]    
	\caption{
		Schematic diagram of the CGAN solver in this study. The color of the initial beam from black to white represents the charge-density from low to high. The compressed temporal profile is obtained by letting the initial beam pass through a chicane having specific \textit{R}$_{56}$, \textit{T}$_{566}$ and \textit{U}$_{5666}$. By feeding the custom temporal profiles and noise components to the trained generator of the CGAN, the CGAN solver is able to predict the required dispersion terms to realize the input target temporal profiles.
		\hfill{}
	}\label{fig1}
\end{figure*}

\section{Methods}\label{II}
In a system of magnetic elements, a transfer map \textit{M} describes the relation of initial conditions $\zeta^\textit{i}$ and final conditions $\zeta^\textit{f}$, which can be symbolically written in the form $\zeta^\textit{f}=\textit{M}\zeta^\textit{i}$. A Taylor map \cite{ref21} that represents the final conditions as a Taylor series of the initial conditions can be described as
\begin{equation}
\label{eq1}
\zeta_j^f=\sum_{k}R_{jk}\zeta_k^i+\sum_{kl}T_{jkl}\zeta_k^i\zeta_l^i+\sum_{klm}U_{jklm}\zeta_k^i\zeta_l^i\zeta_m^i+...
\end{equation}
where \textit{R}, \textit{T} and \textit{U} are the first-, second- and third-order transfer matrices, and \textit{j}, \textit{k}, \textit{l} and \textit{m} are the element indices of the coordinate.
\par
It is empirically found that the CGAN is more effective in dealing with low dimensional problems. Therefore, the transfer map here is needed to be sparse. Fortunately, for beams with small transverse emittance and large energy spread, the contribution of geometrical terms can be neglected and the dispersion terms will tend to dominate. A Taylor map that presents the final longitudinal coordinate \textit{q}$_\textit{z,f}$ of a charged particle as a Taylor series of the initial coordinates can be simplified as
\begin{equation}
\label{eq2}
q_{z,f}= q_{z,0}+R_{56}\delta(q_{z,0})+T_{566}\delta(q_{z,0})^2+U_{5666}\delta(q_{z,0})^3+...
\end{equation}
\begin{equation}
\label{eq3}
\delta(q_{z,0})=h_{1}q_{z,0}+h_{2}q_{z,0}^2+h_{3}q_{z,0}^3+...
\end{equation}
where \textit{q}$_{\textit{z},0}$ is the initial longitudinal coordinate with respect to the bunch center, \textit{R}$_{56}$, \textit{T}$_{566}$ and \textit{U}$_{5666}$ are the first-, second- and third-order longitudinal dispersion terms respectively, $\delta$=$\Delta$\textit{E}/$\textit{E}_{0}$ represents the energy deviation relative to the nominal beam energy, and $h_1$, $h_2$, $h_3$... are the first-, second- and third-order energy chirps respectively. From Eq. (\ref{eq2}), for a initial beam with a specific energy chirp, a desired custom temporal profile can be achieved at the exit of the bunch compressor by using an appropriate combination of longitudinal dispersion terms.\par
In our temporal shaping scheme (see Fig. \ref{fig1}), the first-, second- and third-order longitudinal dispersion terms, i.e. \textit{R}$_{56}$, \textit{T}$_{566}$ and \textit{U}$_{5666}$, labeled with the corresponding temporal profile are taken as training data to train a CGAN. Here we consider the third-order approximation because the third-order longitudinal dispersion term was proven to be important in previous studies and it is rarely necessary to consider even higher order terms for single-pass transport \cite{ref52}. The CGAN used in this study consists of a pair of neural networks, the generator \textit{G} and the discriminator \textit{D}. The generator \textit{G} is trained as a solver that can produce fake samples of longitudinal dispersion terms to realize new temporal profiles. Meanwhile, a discriminator \textit{D} is trained to judge whether the given dispersion terms and temporal profile are drawn from the original training data or is generated by \textit{G}. The \textit{D} will compete with the \textit{G} to force the fake samples generated by \textit{G} to satisfy the distribution learned from the training data. By feeding the trained solver with the target temporal profile and the noise component, the solver is expected to predict different potential solutions to realize the target when multiple solutions exist.\par
In this study, we use an initial electron beam (see Fig. \ref{fig1}) of Gaussian charge-density distribution having an inhomogeneous energy chirp, with the same beam parameters as in Ref. \cite{ref17}. The initial beam is sent to a chicane-type magnetic compressor that have specific \textit{R}$_{56}$, \textit{T}$_{566}$ and \textit{U}$_{5666}$ values, and the custom desired temporal profiles are expected to be obtained at the exit of the chicanes. The layout of the chicane is shown in Fig. \ref{fig1}. The layout of chicane is chosen to such that the \textit{R}$_{56}$, \textit{T}$_{566}$ and \textit{U}$_{5666}$ of this chicane can be adjusted in a large range by tuning strengths of the magnets with a direct search method \cite{ref53}. \par 
To produce training data for the CGAN, 10000 sets of \textit{R}$_{56}$, \textit{T}$_{566}$ and \textit{U}$_{5666}$ samples are stochastically generated within an empirically large enough range. The initial beam is sent to the chicanes that have different stochastic \textit{R}$_{56}$, \textit{T}$_{566}$ and \textit{U}$_{5666}$ values, and the final longitudinal coordinates of the beam at the exit of the chicanes are calculated with an accelerator simulation code \textit{Accelerator Toolbox} \cite{ref54}. The final temporal profile is then obtained and converted to a vector of length 200. The dispersion terms and corresponding temporal profile are treated as feature and label of the real data samples, respectively. A uniform distribution with dimensionality 100 is defined, from which a noise component is randomly selected as the noise vector and fed to \textit{G}.\par
The training of a CGAN is difficult because one has to ensure the balance between the \textit{G} and \textit{D}. When the loss of any one of the \textit{G} and \textit{D} converges quickly to zero while the other doesn't, it will cause the training to fail \cite{ref55}. For this particular case, it is found that the training of \textit{D} is much simpler than the training of \textit{G}. To suppress the training of \textit{D}, the learning rate of \textit{D} is set to be one tenth of that of \textit{G}, i.e. 0.0001 and 0.001 respectively. The training of the networks is implemented with Adam optimizer \cite{ref56} for 10000 epochs on an open source ML library \textit{tensorflow} \cite{ref57}. The training time of the networks is about 20 minutes on a personal computer with a GeForce RTX 2070 Super Graphics Card.\par

\vspace*{1.2mm}
\section{Beam temporal shaping results and discussions}\label{IV}
\vspace*{1.2mm}
\subsection{Realization of two typical temporal profiles}
\noindent
\begin{figure*}[htbp]    
	\vspace*{1mm}    
	\centering    
	\includegraphics[scale=0.4]{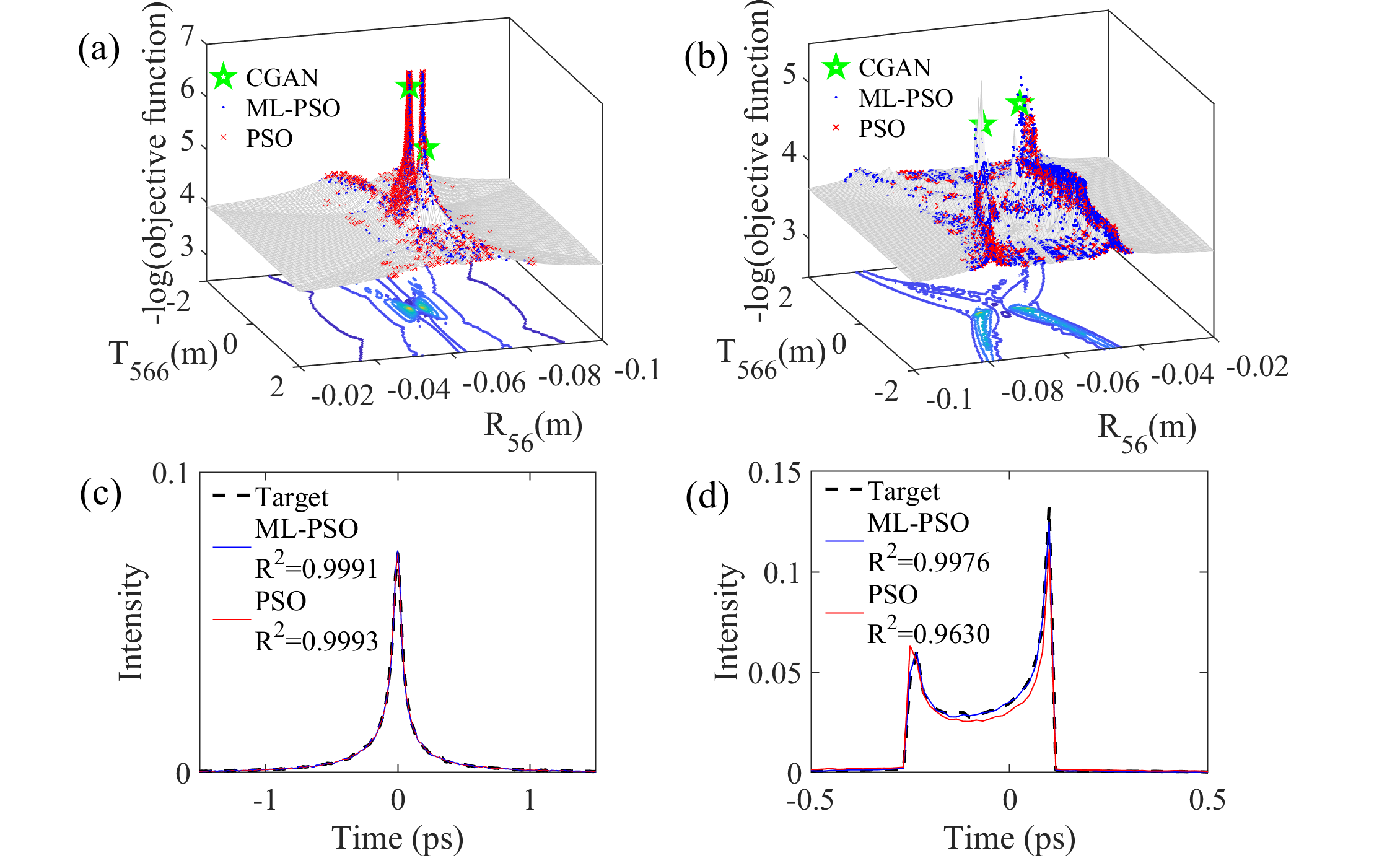}   
	\\[1mm]    
	\caption{
		Using stochastic optimization methods to solve temporal shaping problems. (a) and (b) show the grid scan results and the evolutionary trajectory of the PSO and the surrogate model-based PSO in the variable space for a cusp-shaped profile and a double-horn profile, respectively. The \textit{z}-axis in (a) and (b) is -log(\textit{objective function}) representing the objective performance. The contour maps of the grid scan results are plotted at the bottom, where the color from blue to red represent the objective performance from low to high. For each \textit{R}$_{56}$-\textit{T}$_{566}$ grid in (a) and (b), only one \textit{U}$_{5666}$ with the highest objective performance is plotted. Two separate predictions of the CGAN solver are also shown in (a) and (b) for comparison. (c) and (d) are the final temporal profiles obtained with the optimization methods for the cusp-shaped profile and double-horn profile, respectively. $\textit{R}^2$ is the determination coefficient to the target temporal profile.
		\hfill{}
	}\label{fig2}
\end{figure*}

Two temporal profiles that are common in bunch compression, i.e. a cusp-shaped profile and a double-horn profile (see Fig. \ref{fig2}(c, d)) are used as target profiles to test the performance of our CGAN solver. To look into the details of how the CGAN solver solves the temporal shaping problems, grid scan is first performed in the \textit{R}$_{56}$, \textit{T}$_{566}$ and \textit{U}$_{5666}$ space to find all potential solutions for the two test target profiles, respectively. According to the $100\times100\times100$ grid scan within a empirically large range (see Fig. \ref{fig2}(a, b)), at least two potential solutions that have almost the same highest objective performance are found for each target profiles, which suggests that both temporal shaping problems are one-to-many. Besides the potential solutions, a plenty of local optima are also observed in the variable space of the double-horn profile, indicating that the bunch compression process of the double-horn profile has a stronger nonlinearity.\par
Before implementing the CGAN solver, we first use current state-of-the-art approaches for temporal shaping, the stochastic optimization methods, to realize the two test target profiles. After comparing various optimization methods, the PSO which shows the best performance is used to solve the two test problems. The optimization function is to minimize the mean square error with respect to the target profile, and the free variables in this optimization are the \textit{R}$_{56}$, \textit{T}$_{566}$ and \textit{U}$_{5666}$ of the bunch compressor. All the optimized variables are normalized to a range of [0, 1]. The initial population are 300 solutions randomly generated from a uniform distribution within the variable range. The velocity weight factor is set to be 0.4, and the acceleration coefficients of the group best experience and the personal experience are both set to be 1. In addition, considering the population size can significantly affect the optimization performance, we also try to use a surrogate model-based PSO similar to the method proposed in \cite{ref41} to optimize the test problems with larger population size. In this method, a ML surrogate model is trained to quickly evaluate the candidate solutions that is several times faster than numerical simulation. With the aid of the surrogate model, the population size can be extended to ten times of the standard PSO (i.e. 3000) while the computing time can be remained at the same order of magnitude. To reduce the influence of randomness in the optimization, the optimization is repeatedly performed for five times, and only the best solutions obtained among the repeated tests are selected.\par
The evolutionary trajectories of the optimization are plotted in the variable space (see Fig. \ref{fig2}(a, b)), and the final temporal profiles obtained with the optimization are shown in Fig. \ref{fig2}(c, d). The results in Fig. \ref{fig2}(a) appears that for the cusp-shaped profile, the PSO can find the two potential solutions to realize the target profile with high fitness. While for the double-horn profile that is more complicated, the PSO can only find one potential solution and miss another. However, when the population size is extended to 3000 with the aid of surrogate model, all two potential solutions for the double-horn profile can be actually found. The results indicate that for a one-to-many problem, if the population size is not sufficiently large, the stochastic optimization methods may find one potential solution while with others missed.\par
Then the CGAN solver is implemented to solve the two test problems. The target profiles and the noise components are simultaneously fed to the trained solver which finally results in multiple sets of fake \textit{R}$_{56}$, \textit{T}$_{566}$ and \textit{U}$_{5666}$ samples. It is found that the fake samples converge to several points in the phase space, which represent multiple solutions of a temporal shaping problem. For each of the cusp-shaped profile and double-horn profile, two separate solutions obtained with the CGAN solver are shown in Fig. \ref{fig2}(a, b) in the variable space. For the cusp-shaped profile, one CGAN solution is almost the same as the best solution obtained with grid scan, and another one having slightly lower objective performance is close to a solution obtained with the PSO. For the double-horn profile, the two CGAN predictions are close to the two highest peaks in the grid scan map.\par
\begin{figure*}[htbp]    
	\vspace*{1mm}    
	\centering    
	\includegraphics[scale=0.4]{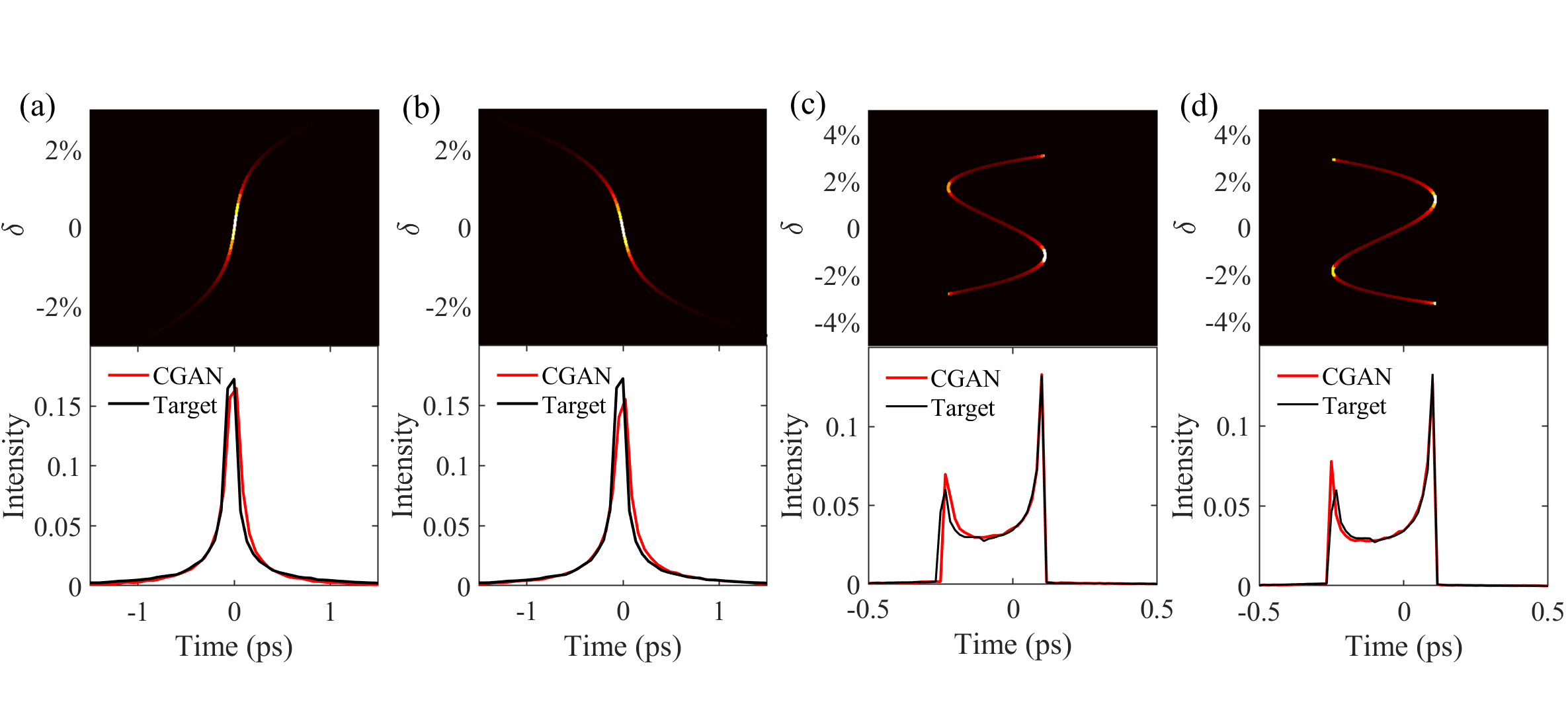}   
	\\[1mm]    
	\caption{
		Longitudinal phase space distribution and temporal profiles of two separate CGAN predictions. The left two columns represent the cusp-shaped profile, and the right two columns represent the double-horn profile, respectively.
		\hfill{}
	}\label{fig3}
\end{figure*}
The longitudinal phase space distribution resulted from the CGAN predictions are shown in Fig. \ref{fig3}. It is found that the beams in Fig. \ref{fig3}(b, d) are over compressed, i.e. the head and tail of the beam are reversed. The over compressed beam finally results in almost the same temporal profiles as the under compressed beam in Fig. \ref{fig3}(a, c), with a high determination coefficient close to 1.\par
The results in Fig. \ref{fig2} and Fig. \ref{fig3} indicate that the CGAN solver can predict the longitudinal dispersion terms to realize the custom desire temporal profiles with high accuracy. Furthermore, the CGAN solver is able to give different solutions for the same input temporal profile when multiple solutions exist, which may be a limitation of stochastic optimization methods that sometimes lead to one solution with insufficiently large population size. The acquirement of the multiple solutions in Fig. \ref{fig3} is crucial for temporal shaping since an under compressed beam and an over compressed beam can provide different benefits in scientific applications. For example, an over compressed scheme has the potential to provide larger bandwidth FEL radiation \cite{ref23}. However, compared to the under compressed beam, the over compressed beam also leads to a significant coherent synchrotron radiation (CSR) that can reduce the slice alignment and spoil the transverse emittance of the electron beam \cite{ref22}. Besides, it is possible to select one from the obtained multiple solutions that is more feasible to implement in practical scientific experiments by experienced operators or an additional evaluator. For instance, it is found that to realize the cusp-shaped profile, the octupole strength required to achieve the longitudinal dispersion terms of one obtained solution is significantly higher than that of another solution ($-15.7$m$^{-3}$ and $0.4$m$^{-3}$ respectively). The strong octupoles are not trivial to build and will bring high sensitivity during the beamline optimization and operation.\par
\begin{figure}[t]
	\centering
	\includegraphics[scale=0.4]{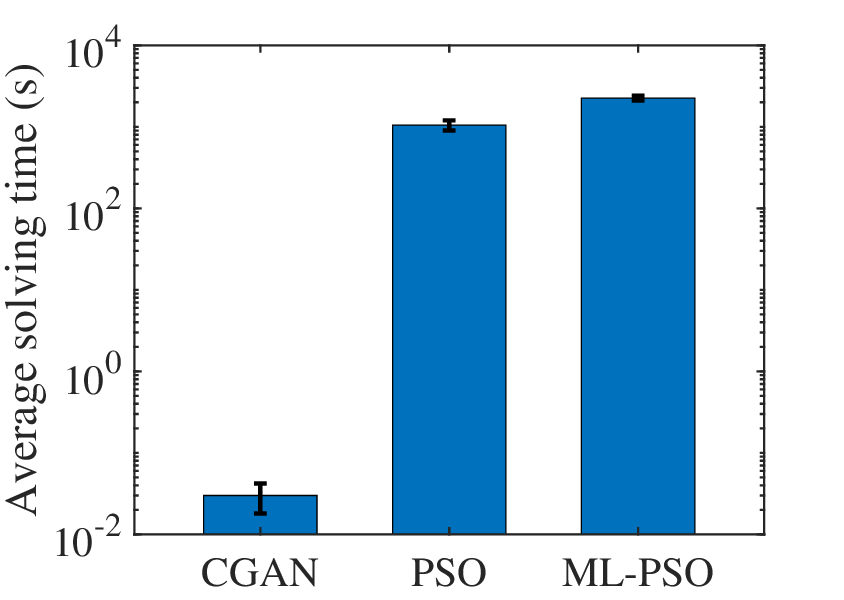}
	\caption{The average solving time with the CGAN solver, the PSO and the surrogate model-based PSO for five repeated tests.}
	\label{fig4}
\end{figure}
Fig. \ref{fig4} shows the average solving time for different methods. It is noted that for any target profile, it takes dozens of minutes to perform the optimization with the stochastic optimization methods. The optimization results are not reusable for a new target profile. This optimization time can be much longer when some complicated effects like space charge and CSR are considered. Compared to the stochastic optimization methods, once the CGAN is trained, it only needs little time (fractions of one second) to directly predict the dispersion terms for any new temporal profile, which can be several orders of magnitude faster than using stochastic optimization methods.\par
\subsection{Consideration of CSR effect}
When a high energy electron beam is compressed to be denser and shorter, CSR effect will be significant and play an important role in the beam dynamics. To explore the effectiveness of the CGAN solver to shape beam profile under strong CSR effect, the dispersion terms and corresponding lattices settings used above are converted to ELEGANT \cite{ref58} lattice files with 1D CSR effect taken into account. The charge of the beam is set to be 500pC and the central energy is 1GeV. It takes several hours to generate about 10000 data samples on a personal workstation. Similar to above works, we train a new CGAN (with the same hyperparameter settings) with the new tracking results as training data. \par
The same target profiles, i.e., the cusp-shape and double-horn profiles, are used to test the performance of the CGAN solver. As Fig. 5 shows, for the cusp-shape profile, the beam profile obtained with the CGAN solution generally matches the target profile, which is slightly wider than the target. For the double-horn profile, the head of the beam matches the target profile but a mismatch is observed at the tail of the beam. It may be because that the target profiles are generated by non-CSR tracking and it may be difficult to produce the same beam profile when CSR effect is considered. To our experience of using CGAN, when an unattainable target is given as the input label, the CGAN can generate samples that match the target as closely as possible, like the results in Fig. 5. The tolerance of CGAN to the realizability of the given target might vary from case to case and is challenging to measure since the neural network itself is a "black box". Considering this challenge, in an explorative research, the stochastic optimization methods like PSO are still a general solution that cannot be simply replaced by the CGAN solver. In the application of the CGAN solver, if possible, an expert in the applied domain is recommended to ensure that the given targets are not too far from reality.

\begin{figure}[t]
	\centering
	\includegraphics[scale=0.35]{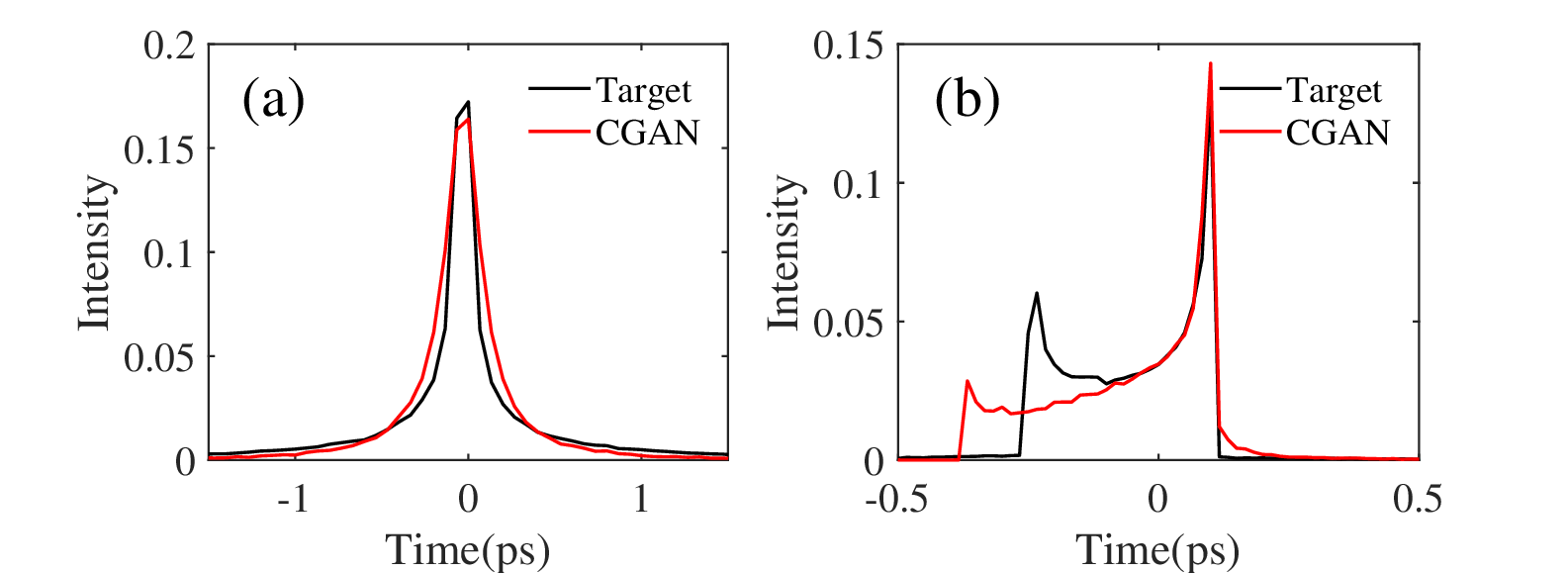}
	\caption{Beam profiles obtained with CGAN solutions for two target profiles with CSR effect taken into account. (a) and (b) represent a cusp-shape target and a double-horn target, respectively.}
	\label{fig5}
\end{figure}
\subsection{Realization of two temporal profiles with scientific merits}
\begin{figure}[t]
	\centering
	\includegraphics[scale=0.27]{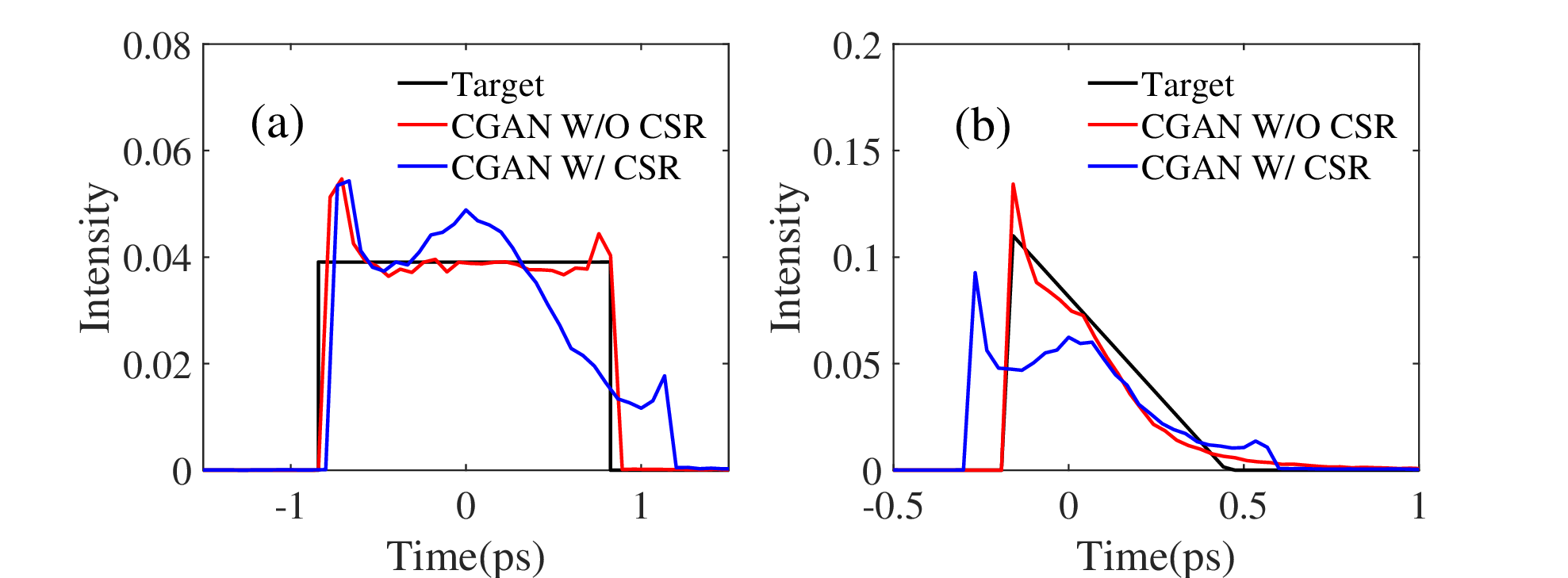}
	\caption{The temporal profiles obtained with the CGAN solver for the flat-top profile (left) and the linearly-ramped profile (right). The red/blue solid line represents the results are predicted by the CGAN model without/with CSR effect considered in the training.}
	\label{fig6}
\end{figure}

In addition to the above two common temporal profiles, two other temporal profiles, namely the flat-top profile and the linearly-ramped profile (see Fig. \ref{fig6}), are also studied. These two additional temporal profiles are more frequently-used in various scientific applications and have greater scientific merit. The flat-top temporal profile is desired in FELs to reduce the third-order curvature in the time-energy correlation due to wakefield, so as to obtain better FEL performance with improved pulse energy, peak power and bandwidth control \cite{ref16}. The linearly-ramped temporal profile is treated as the optimal shape of the drive beam in PWFAs to supply a high transformer ratio because it maximizes the energy that can be gained by a trailing particle accelerated in its wakefield \cite{ref14}.\par
The two additional test temporal profiles are fed to the same trained CGAN solvers (both CGAN solvers with/without CSR effect considered are used separately) to predict multiple fake samples of longitudinal dispersion terms. The final temporal profiles resulted from the CGAN solutions are illustrated in Fig. \ref{fig5}. When CSR effect is not considered, the CGAN solver can predict the longitudinal dispersion terms with high accuracy, resulting in almost the same temporal profiles compared to the target profile, especially for the linearly-ramped profile. For the flat-top profile, horns occur at the head and the tail of the compressed beam, which cannot be completely eliminated due to the nature of bunch compression with a single chicane compressor. The horns may be further flattened with an additional bunch compressor \cite{ref59} that is, however, beyond the scope of this study. Nevertheless, the FWHM of the horns is very narrow compared with the flat part of the bunch when CSR effect is not considered. When CSR effect is taken into account, the CGAN predictions can result in compressed beams with close duration as the target profiles. However, it seems difficult to shape the beam to the same regular shapes as the target profiles with the CGAN solver. One explanation might be that under strong CSR effect, the beam itself cannot form these regular target profiles after travelling through such a simple chicane compressor. Nevertheless, if necessary, the CGAN solutions can be also used as good initial solutions for further exploration of the potential solutions for the target profiles.\par

\vspace*{-0.2mm}

\section{Conclusion}\label{VII}

\noindent
We have proposed a CGAN solver for one-to-many problems of temporal shaping of electron beams. By learning from the stochastically generated data, a trained CGAN solver can quickly and accurately predict available combinations of the dispersion terms up to the 3rd order to realize desired custom temporal profiles. For one-to-many problems, the CGAN can predict many, if not all the, potential solutions simultaneously by receiving different input noise vectors. Once the CGAN is trained, it can produce the required dispersion terms for a new target profile within fractions of a second, showing orders of magnitudes faster computing speed than the conventional method, i.e., the stochastic optimization methods. \par 
This method can be easily transferable to other similar problems, for instance, photon pulse shaping and transverse phase space manipulation of an electron bunch. We expect that the CGAN solver can serve as a direct and real-time method to solve one-to-many problems in more scientific applications.\par

\acknowledgements{%
	The authors thank Dr. Juhao Wu for nice discussion and suggestions. This work is supported by National Natural Science Foundation of China (No. 11922512), Youth Innovation Promotion Association of Chinese Academy of Sciences (No. Y201904) and National Key R\&D Program of China (No. 2016YFA0401900). 
}

\begin{small}
	
\end{small}

\end{document}